\documentclass[aps,prd,preprint,superscriptaddress,tightenlines,nofootinbib]{revtex4}

\usepackage{graphicx}
\usepackage{dcolumn}
\usepackage{bm}
\usepackage{epsfig}

\def\mus{$m_b^{\rm 1S}$}
\def\epm{$e^+e^-$}

\def\elep{$e^{\pm}$}
\def\mulep{$\mu^{\pm}$}
\newcommand{\nc}{\newcommand}
\newcommand{\rnc}{\renewcommand}
\nc{\bea}{\begin{eqnarray}}
\nc{\eea}{\end{eqnarray}}
\rnc{\thefootnote}{\fnsymbol{footnote}}

\begin{document}

\preprint{\tighten\vbox{\hbox{\hfil CLNS 02/1810}
                        \hbox{\hfil CLEO 02-16}}}

\title{Measurement of lepton momentum moments in the decay
$\bar{B} \rightarrow X \ell {\bar \nu}$ and determination of
Heavy Quark Expansion parameters and $| V_{cb}| $.}



\author{A.~H.~Mahmood}
\affiliation{University of Texas - Pan American, Edinburg, Texas 78539}
\author{S.~E.~Csorna}
\author{I.~Danko}
\affiliation{Vanderbilt University, Nashville, Tennessee 37235}
\author{G.~Bonvicini}
\author{D.~Cinabro}
\author{M.~Dubrovin}
\author{S.~McGee}
\affiliation{Wayne State University, Detroit, Michigan 48202}
\author{A.~Bornheim}
\author{E.~Lipeles}
\author{S.~P.~Pappas}
\author{A.~Shapiro}
\author{W.~M.~Sun}
\author{A.~J.~Weinstein}
\affiliation{California Institute of Technology, Pasadena, California 91125}
% ...
\collaboration{CLEO Collaboration} 
\noaffiliation

\date{\today}

\setcounter{footnote}{0}

\begin{abstract}
We measure the primary lepton momentum spectrum in $\bar{B}
\rightarrow X \ell {\bar \nu}$ decays, for p$_{\ell}$ $\ge$ 1.5 GeV/$c$
in the $B$ rest frame. From this, we calculate various moments of
the spectrum. In particular, we find 
$R_0\equiv \int _{1.7\ \rm  GeV} (d\Gamma/dE_{sl})dE_{\ell}/\int _{1.5\ \rm GeV} 
(d\Gamma/dE_{sl})dE_{\ell}=0.6187\pm 0.0014_{stat}\pm 0.0016_{sys}$ and
$R_1\equiv 
\int _{1.5\ \rm GeV} E_{\ell}(d\Gamma/dE_{sl})dE_{\ell}/\int _{1.5\ \rm GeV} (d\Gamma/dE_{sl})dE_{\ell}=(1.7810\pm 0.0007_{stat} \pm 0.0009_{sys})$ GeV.
We use these moments to determine non-perturbative 
parameters governing the semileptonic width. In particular, we
extract the Heavy Quark Expansion  parameters
$\bar{\Lambda}=(0.39\pm 0.03_{stat}\pm 0.06_{sys}\pm 0.12_{th}$)
GeV and $\lambda _1=( -0.25\pm 0.02_{stat}\pm 0.05_{sys}\pm
0.14_{th}$) GeV$^2$. The theoretical constraints used are
evaluated  through order $1/M_B^3$ in the non-perturbative
expansion and $\beta_0\alpha_s^2$ in the perturbative expansion.
We use these parameters to extract $|V_{cb}|$ from the world
average of the semileptonic width and find $|V_{cb}|=(40.8\pm
0.5_{\Gamma _{sl}}\pm 0.4_{(\lambda _1,\bar{\Lambda})_{exp}}\pm
0.9_{th}) \times 10^{-3}$. In addition, we extract the short
range $b$-quark mass \mus\ = ($4.82\pm 0.07_{\exp}\pm
0.11_{th}$) GeV/$c^2$. Finally, we discuss the implications of
our measurements for the theoretical understanding of inclusive
semileptonic processes.
\end{abstract}

\maketitle
\newpage

\section{Introduction}
Experimental data on inclusive $B$ meson semileptonic decays can in
principle provide a very precise method to determine the
Cabibbo-Kobayashi-Maskawa (CKM) quark mixing parameter $|V_{cb}|$ \cite{pdg2002}. 
A crucial theoretical input is the hadronic matrix element
needed to express the measured semileptonic width in terms of
$|V_{cb}|$. The Heavy Quark Expansion (HQE)
\cite{manohar-wise,bigi,gremm-kap,secondalph} is a QCD-based
approach to inclusive processes that casts perturbative and
non-perturbative corrections to the partonic width as power series
expansions. An underlying assumption of this approach is
quark-hadron duality. It is important to quantify the
uncertainties induced by the neglected higher order terms in the
non-perturbative expansion, as well as the uncertainty introduced
by possible duality violations, in order to achieve a full
understanding of the theoretical errors and be able to ascertain
the true uncertainty on $|V_{cb}|$. The only strategy proposed so
far to gather further insight is to measure several quantities
predicted in this framework. A precise measurement of the lepton
spectrum is an important element of this program and is the key
result presented in this paper.

The theoretical expression for the inclusive semileptonic width
for ${\bar B} \rightarrow X \ell {\bar \nu}$ ($\ell=\mu$ or $e$)
through ${\cal O}(1/M_B^3)$ in the non-perturbative expansion and
$\beta_0 ({\alpha_s}/{\pi})^2$ in the perturbative one is
given by \cite{gremm-kap,vcb1}
 \bea \Gamma_{sl} &=& \frac{{G_F}^2
|{\mbox{V}}_{cb}|^2 {M_B}^{5}}{192 \pi^3} \; 0.3689 \left[ 1-1.54
\frac{\alpha_s}{\pi} -
1.43 \beta_0 \left(\frac{\alpha_s}{\pi}\right)^2 \right. \nonumber \\
 & & - 1.648 \frac{\bar{\Lambda}}{{M_B}} \left(1-0.87 \frac{\alpha_s}{\pi}\right)
- 0.946 \left(\frac{\bar{\Lambda}}{{M_B}}\right)^2
-3.185 \frac{{\lambda}_{1}}{{M_B}^{2}}+0.02\frac{{\lambda}_{2}}{{M_B}^{2}} \nonumber \\
 & &-0.298\left(\frac{\bar{\Lambda}}{{M_B}}\right)^3
-3.28 \frac{{\lambda}_{1} \bar{\Lambda}}{{M_B}^{3}}
 + 10.47 \frac{{\lambda}_{2} \bar{\Lambda}}{{M_B}^{3}}
- 6.153 \frac{{\rho}_{1}}{{M_B}^{3}}+ 7.482 \frac{{\rho}_{2}}{{M_B}^{3}}\nonumber \\
& & \left.- 7.4 \frac{{\tau}_{1}}{{M_B}^{3}} + 1.491 \frac{{\tau}_{2}}{{M_B}^{3}}
 - 10.41 \frac{{\tau}_{3}}{{M_B}^{3}}
 - 7.482 \frac{{\tau}_{4}}{{M_B}^{3}}+ {\cal O}\left(\frac{1}{{M_B}^{4}}\right)
 \right],
\label{vcbf} \eea where $\beta_0 = (33 - 2 n_f)/3 = 25/3$ is the
one-loop QCD beta function and $n_f$ is the number of relevant flavors
and the form factors $\rho_1$,
$\rho_2$, $\tau_1$, $\tau_2$, $\tau_3$, and $\tau_4$ are the
parameters of the $1/M_B^3$ terms in the non-perturbative
expansion. These $1/M_B^3$ form factors are expected, from
dimensional arguments, to be of the order $\Lambda _{QCD}^3$, and
thus they are generally assumed to be $\leq (0.5)^{3}$ GeV$^3$. In
addition, $\rho _1$ is expected to be positive from the
vacuum-saturation approximation \cite{ikaros-rho1}. Furthermore,
as Gremm and Kapustin have noted \cite{gremm-kap}, the $B^{\star}$-$B$
and $D^{\star}$-$D$ mass splittings impose the constraint
\begin{equation}
\rho _2-\tau _2 -\tau _4= \frac{\kappa(m_c)M_B^2 \Delta M_B (M_D +
\bar{\Lambda})- M_D^2\Delta M_D (m_B+\bar{\Lambda})}
{M_B+\bar{\Lambda} - \kappa(m_c) (M_D+\bar{\Lambda})},
\end{equation}
where $m_b$ and $m_c$ represent the beauty and charm quark masses, respectively;
$\kappa(m_c)\equiv [\alpha _s(m_c)/\alpha
_s(m_b)]^{(3/\beta_0)}$ and $\Delta M_B (\Delta M_D)$ represents the
vector-pseudoscalar meson splitting in the beauty (charm)
sector.

The parameter ${{\lambda}_{1}}$ \cite{manohar-wise,bigi} is
related to the expectation value of the operator corresponding to  the kinetic 
energy of the $b$ quark inside the $B$
meson:
\begin{equation}
{{\lambda}_{1}} =\frac{1}{2 M_B} \left< B(v)|{\bar{h}_v} (i D)^2 h_v|B(v)\right>,
\end{equation}
\noindent where $v$ denotes the 4-velocity of the heavy hadron and ${h}_v$ is 
the quark field in
the heavy quark effective theory.
The parameter ${{\lambda}_{2}}$ \cite{manohar-wise,bigi} is the
expectation value of the leading chromomagnetic operator that breaks the heavy
quark spin symmetry. It is formally defined as
\begin{equation}
{{\lambda}_{2}} =\frac{-1}{2 M_B} \left<B(v)|{\bar{h}_v} \frac{g}{2} \cdot \sigma^{\mu\nu}
G_{\mu\nu} h_v|B(v)\right>,
\end{equation}
where $h_v$ is the heavy quark field and $\left|B(v)\right>$ is the
$B$ meson
state. The value of $\lambda _2$ is determined from the $B^{\star}-B$ mass difference
to be $0.128\pm 0.010$ GeV$^2$. The quantity ${\bar{\Lambda}}$ is related to
the $b$-quark pole mass $m_b$ \cite{manohar-wise,bigi} through the
expression
\begin{equation}
m_b = {\bar{M}_B}-{\bar{\Lambda}}+\frac{{\lambda}_{1}}{2 m_b},
\end{equation}
where ${\bar{M}_B}$ is the spin-averaged $B^{(\star)}$ mass
(${\bar{M}_B} = 5.313$ GeV/$c^2$). A similar relationship holds between
the $c$-quark mass $m_c$ and the spin-averaged charm meson mass
(${\bar{M}_D} = 1.975$ GeV/$c^2$).

The shape of the lepton momentum spectrum in ${\bar B} \rightarrow X \ell {\bar \nu}$ decays
can be used to measure the HQE parameters ${{\lambda}_{1}}$
and ${\bar{\Lambda}}$, through its energy moments, which are also predicted in
the Heavy Quark Expansion. We choose to study truncated moments of the lepton
spectrum, with a momentum cut of $p_{\ell} \ge 1.5$
GeV/$c$ in the $B$ meson rest frame. This choice decreases the
sensitivity of our measurement to the secondary leptons from the cascade
decays ($b \rightarrow c \rightarrow s \ell \nu\ {\rm or}\ d \ell\nu$).

We extract the HQE parameters $\bar{\Lambda }$ and $\lambda _1$
from measurements of two moments originally suggested by Gremm
{\it et al.} \footnote{Our notation is different
than that used in Ref.~\protect{\cite{ligeti}}, where R$_0$ is
first introduced as R$_2$.}:
\begin{equation}
{\mbox{R}}_{0} =\frac{\int_{1.7\ \rm GeV}^{} (d \Gamma_{sl}/dE_{\ell})
dE_{\ell}}{\int_{1.5\ \rm GeV}^{} (d \Gamma_{sl}/dE_{\ell}) dE_{\ell}} \ {\rm and}
\label{r2}
\end{equation}

\begin{equation}
{\mbox{R}}_{1} =\frac{\int_{1.5\ \rm GeV}^{} E_{\ell} (d \Gamma_{sl}/dE_{\ell})
dE_{\ell}}{\int_{1.5\ \rm GeV}^{} (d {\Gamma_{sl}} /dE_{\ell}) dE_{\ell}}.
\label{r1}
\end{equation}
The theoretical expressions for these moments R$^{th}_{0,1}$
\cite{gremm-kap,chris} are evaluated by
integrating the dominant  $b\rightarrow c
\ell \bar{\nu}$ component of the lepton spectrum. In
addition, the small contribution coming from charmless
semileptonic decays $b\rightarrow u \ell \bar{\nu}$ is included
 by adding the contribution from $d\Gamma
_{u}/dE_{\ell}$, scaled by
$|V_{ub}/V_{cb}|^2$ \cite{ligeti,chris}.

We determine these two moments from the measured lepton
spectrum in $\bar{B} \rightarrow X \ell \bar{\nu}$ and insert them
in the theoretical expressions to extract the two parameters
$\lambda_1$ and $\bar{\Lambda}$.  We have previously published
experimental determinations of $\bar{\Lambda}$ and $\lambda _1$
obtained by studying the $E_\gamma$ spectrum in $b\rightarrow s
\gamma$ \cite{bsgamma} and the first moment of the mass $M_X$ of
the hadronic system recoiling against the $\ell\bar{\nu}$ pair in
$\bar{B}\rightarrow X \ell \bar{\nu}$ decays~\cite{eht-moments}.
We compare our results to these measurements.

In recent years, increasing attention has been focused on ``short-%
range masses,'' preferred by some authors as they are not affected
by renormalon ambiguities \cite{ikaros-kolya}.  In particular,
the so-called 1S $b$-quark mass, \mus, defined as one half of the energy 
of the 1S $b\bar{b}$ state calculated in perturbation theory, 
has been extracted from
$\Upsilon$(1S) resonance data \cite{h1}. The mass \mus\ has been shown to have remarkably well-behaved perturbative relations to other physical quantities
such as the hadronic matrix element governing the $b\rightarrow u$ semileptonic width \cite{aida:mike}. Using the formalism
developed by Bauer and Trott \cite{chris}, we have used
the spectral moments $R_0$ and $R_1$ to determine \mus .

These authors also explore different lepton energy moments, by
varying the exponent of the energy in the integrands and the lower
limits of integration. In particular, they identify several
moments that provide constraints for
 \mus\ and $\lambda_1$ that are less sensitive to higher
order terms in the non-perturbative expansion. We study four such
moments defined as

\begin{equation}
{\mbox{R}}^{(3)}_{a} =\frac{\int_{1.7\ \rm GeV}^{} E_{\ell}^{0.7}(d \Gamma_{sl}/dE_{\ell})
dE_{\ell}}{\int_{1.5\ \rm GeV}^{} E_{\ell}^2 (d \Gamma_{sl}/dE_{\ell}) dE_{\ell}},\
\label{r3a}
\end{equation}
\begin{equation}
{\mbox{R}}^{(3)}_{b} =\frac{\int_{1.6\ \rm GeV}^{} E_{\ell}^{0.9}(d \Gamma_{sl}/dE_{\ell})
dE_{\ell}}{\int_{1.7\ \rm GeV}^{} (d \Gamma_{sl}/dE_{\ell}) dE_{\ell}} \label{r3b},
\end{equation}
\begin{equation}
{\mbox{R}}^{(4)}_{a} =\frac{\int_{1.6\ \rm GeV}^{}E_{\ell}^{0.8} (d \Gamma_{sl}/dE_{\ell})
dE_{\ell}}{\int_{1.7\ \rm GeV}^{} (d \Gamma_{sl}/dE_{\ell}) dE_{\ell}},\ {\rm and}
\label{r4a}
\end{equation}
\begin{equation}
{\mbox{R}}^{(4)}_{b} =\frac{\int_{1.6\ \rm GeV}^{} E_{\ell}^{2.5} (d
\Gamma_{sl}/dE_{\ell}) dE_{\ell}}{\int_{1.5\ \rm GeV}^{}E_{\ell}^{2.9} (d {\Gamma_{sl}}
/dE_{\ell}) dE_{\ell}}. \label{r4b}
\end{equation}
The values of $\bar{\Lambda}$ and $\lambda _1$ determined 
with the latter set of constraints have different relative 
weights of the experimental and theoretical uncertainties and thus provide
complementary information.

Finally, Bauer and Trott identify moments that are insensitive to
\mus\ and $\lambda_1$. They suggest that a comparison between 
theoretical evaluations of these ``duality moments'' and their
experimental values may provide useful constraints on possible
quark-hadron duality violations in semileptonic processes. We
report our measurement of two such ``duality moments", defined as
\begin{equation}
{\mbox{D}}_{3} =\frac{\int_{1.6\ \rm GeV}^{} E_{\ell}^{0.7} (d \Gamma_{sl}/dE_{\ell})
dE_{\ell}}{\int_{1.5\ \rm GeV}^{} E_{\ell}^{1.5}(d {\Gamma_{sl}} /dE_{\ell}) dE_{\ell}}
\label{d3}
\end{equation}
and
\begin{equation}
{\mbox{D}}_{4} =\frac{\int_{1.6\ \rm GeV}^{} E_{\ell}^{2.3} (d \Gamma_{sl}/dE_{\ell})
dE_{\ell}}{\int_{1.5\ \rm GeV}^{}E_{\ell}^{2.9} (d {\Gamma_{sl}} /dE_{\ell}) dE_{\ell}}.
\label{d4}
\end{equation}
This measurement, together with new emerging experimental
information \cite{babarmom,lepmom}, may eventually lead to a more
complete assessment of our present understanding of inclusive
semileptonic decays.

\section{Experimental Method}
The data sample used in this study was collected with the CLEO II
detector \cite{cleoii} at the CESR~ \epm\ collider. It
consists of an integrated luminosity of 3.14 fb$^{-1}$ at the
$\Upsilon$(4S) energy, corresponding to a sample of $3.3\times 10^{6}$
$B\bar B$ events. The continuum background is studied with a sample of
$1.61$ fb$^{-1}$ collected at an energy about $60$ MeV below the
resonance.

We measure the momentum spectrum of electrons and muons with a
minimum momentum of 1.5 GeV/$c$ in the $B$ meson center-of-mass
frame. This momentum requirement ensures good efficiency and
background rejection for both lepton species, thereby allowing us
to check systematic effects with the $\mu/e$ ratio. For muons we
have adequate efficiency and background rejection only above $p_{\mu} \sim$ 
1.3 GeV/$c$. In addition, in this range the inclusive spectra are dominated by the
direct $b\rightarrow c\ell \nu$ semileptonic decay, with only a
small contamination by secondary leptons produced in the decay
chain $b\rightarrow c \rightarrow (s \ell \nu\ {\rm or}\ d\ell \nu)$.

Electrons are identified with a likelihood method that includes several
discriminating variables, most importantly the ratio $E/p$ of the
energy deposited in the electromagnetic calorimeter to the
reconstructed momentum, and the specific
ionization in the central drift chamber.  Muon candidates are required
to penetrate at least five nuclear interaction lengths of absorber material.
 We use the central part of the detector
($|\cos{\theta}| \le 0.71$ for electrons and $|\cos{\theta}| \le 0.61$
for muons).

The overall efficiency is the product of three factors: the
reconstruction efficiency, including event selection criteria and
acceptance corrections; the tracking efficiency; and the $\mu$ or
$e$ identification efficiency. The first two factors are estimated
with Monte Carlo simulations and checked with data, whereas the
lepton identification efficiencies are studied with data:
radiative $\mu$-pair events for the $\mu$ efficiency and radiative
Bhabha electron tracks embedded in hadronic events for the $e$
efficiency. The $e$ identification efficiency is nearly constant
in our momentum range and equal to $(93.8 \pm 2.6) \%$.  The $\mu$
momentum threshold is near our low momentum cut, and the
efficiency rises to a plateau of about 95\% above 2.0 GeV/$c$. 
The distortion in momentum induced by radiation emitted in
the detector and other instrumental effects is corrected for by 
using the same Monte 
Carlo samples used in the efficiency correction.

\begin{figure}[htbp]
\center{ {\epsfig{figure=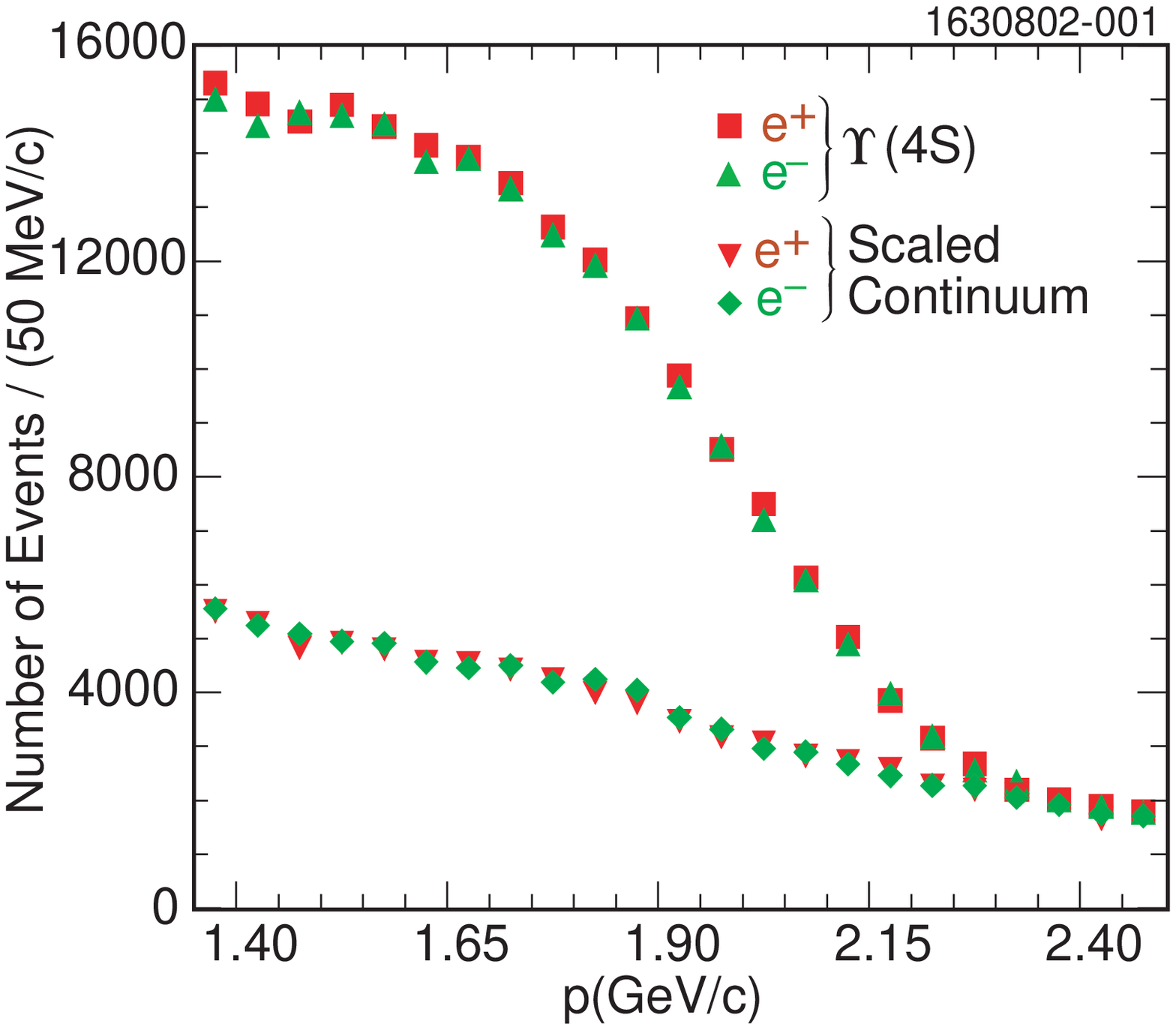,width=4in,height=3.5in}}
{\epsfig{figure=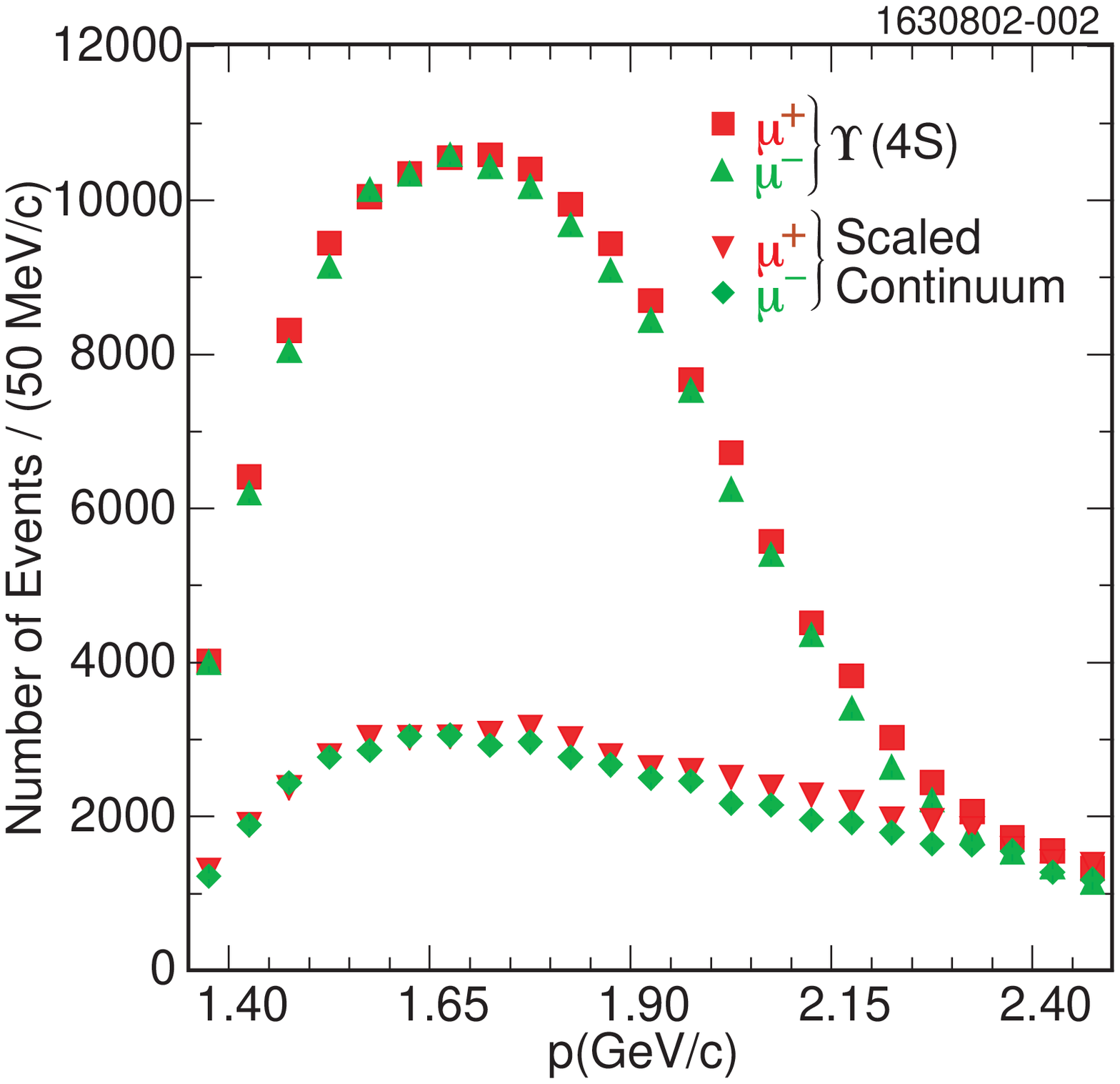,width=4in,height=3.5in}} }
\caption{\small {Raw lepton spectra from the 
$\Upsilon(4S)$ and scaled continuum.}}
\label{totelspec}
\end{figure}

Figure \ref{totelspec} shows the raw yields for electrons (top)
and muons (bottom) from the $\Upsilon$(4S) sample and the
continuum background. The latter is estimated from scaled off-resonance data. The
scaling factor for the continuum sample is determined by the ratio
of integrated luminosities and continuum cross sections and is
$1.930\pm 0.013$. This scale factor has been determined
independently using tracks with momenta higher than the kinematic
limit for $B$-meson decay products. In all the cases no statistically significant lepton
yield has been observed beyond the endpoint for $B$ decays, within
errors. The study of these control samples is used to determine
the systematic error on the continuum scale factor.

The raw yields include hadrons misidentified as leptons (fakes).
This contribution is determined from data as follows. Fake rates
are determined from tagged samples: charged pions from
$K^0_S\rightarrow \pi^+\pi^-$, charged kaons from $D^{\star +}
\rightarrow D^0 \pi^+, D^0 \rightarrow K^-\pi^+$, and protons from
$\Lambda \rightarrow p \pi^-$. The momentum-dependent
probability for misidentifying a hadron track as an electron or muon is
then determined by weighting the pion, kaon, and proton probabilities
according to particle abundances determined with $B\bar{B}$ Monte Carlo. 
The fake correction applied to the data is obtained by folding these fake
probabilities with the measured spectra of hadronic tracks in $B\bar{B}$
events.

We correct for several sources of real leptons. Leptons from
$J/\psi$ decays are vetoed by combining a candidate with another
lepton of the same type and opposite sign and removing it if their
invariant mass is within $3\sigma$ of the known $J/\psi$ mass. A
correction is made for veto inefficiency. A similar procedure is
applied to electrons and positrons coming from $\pi^0$ Dalitz
decays and from $\gamma$ conversions.

\begin{figure}[htbp]
\center{
{\epsfig{figure=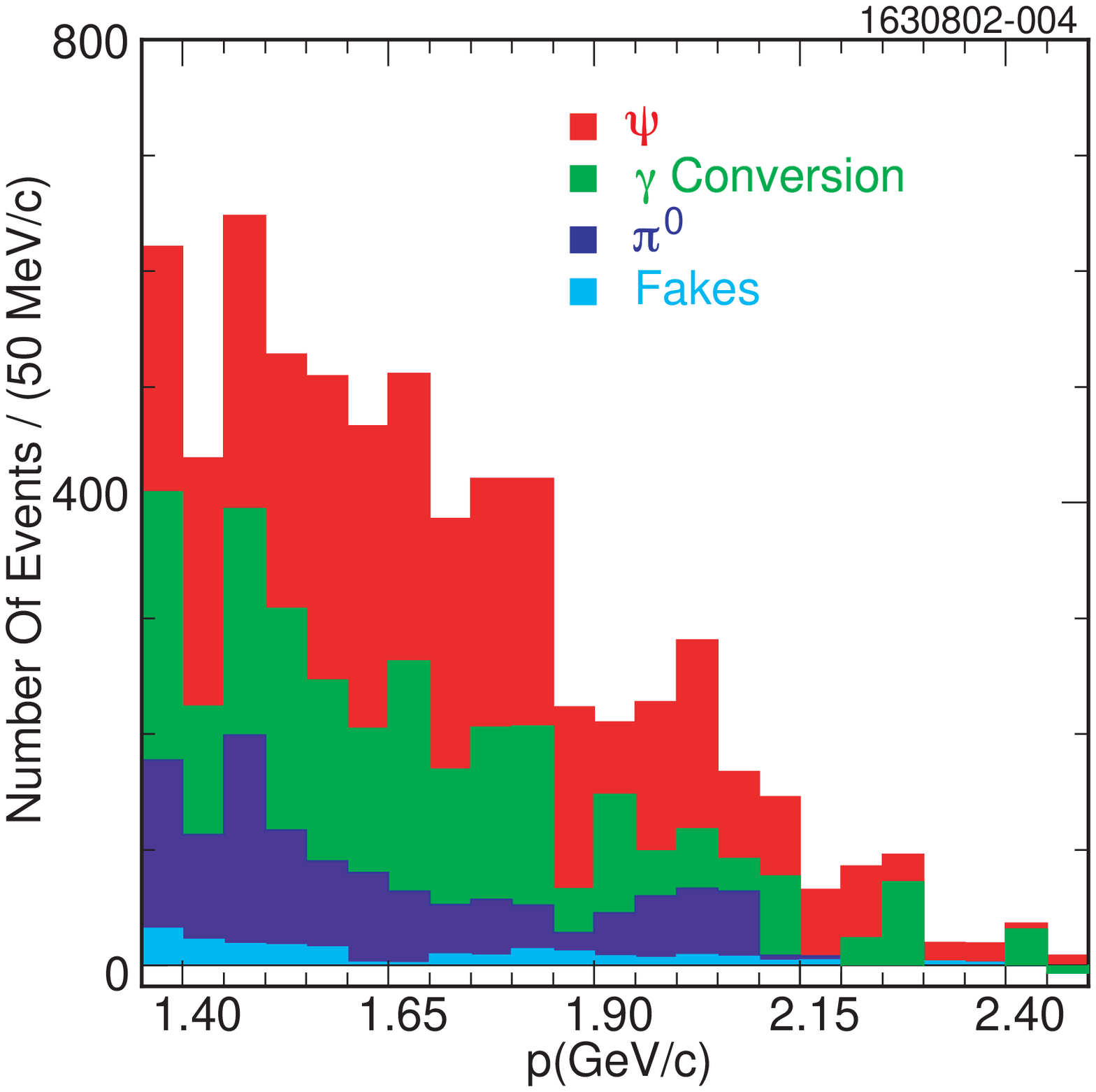,width=4in,height=3.5in}}
{\epsfig{figure=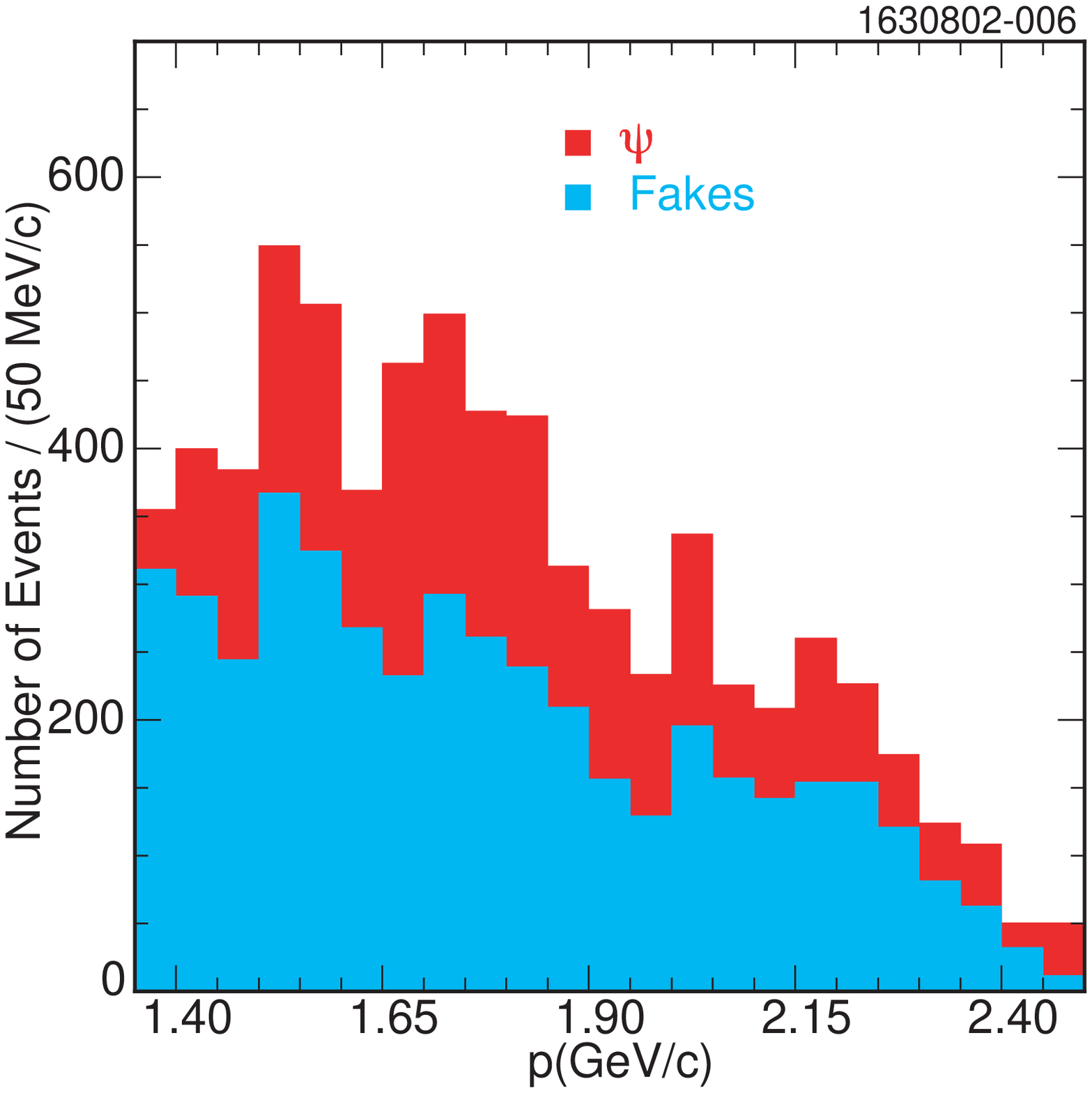,width=4in,height=3.5in}}
}
\caption{\small {
Background components of the electron (top) and muon (bottom) spectra from
processes that are estimated with data.
}}
\label{elbkgcntri}
\end{figure}

\begin{figure}[htbp]
\center{
{\epsfig{figure=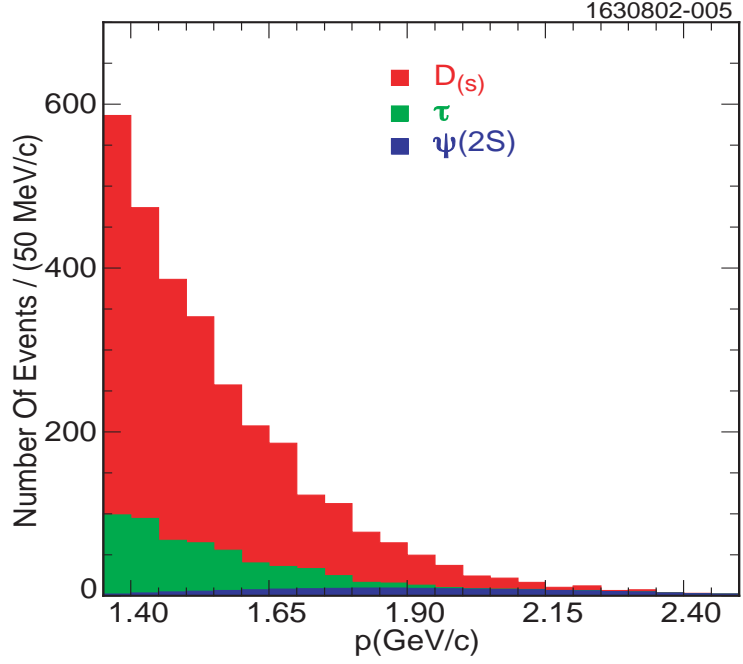,width=4in,height=3.5in}}
}
\caption{\small {Background components of the electron spectrum
that are studied with Monte Carlo simulations; these components are similar in the muon case.}}
\label{elsec}
\end{figure}

Finally, we subtract leptons coming from $\psi(2 \rm S)$ decays or the
secondary decays  $b\rightarrow c\rightarrow (s \ell \nu\ {\rm or}\ d\ell \nu)$ and
$B\rightarrow \tau \rightarrow \ell \nu \bar{\nu}$
using Monte Carlo simulations. Figures \ref{elbkgcntri} and \ref{elsec}
show the individual estimated background contributions to our sample.
Note that all of the backgrounds are small compared to the signal.

Our goal is a precise determination of the shape of the lepton
momentum spectrum, so corrections for the distortion introduced by
electroweak radiative effects are important. We use the prescription
developed by D.~Atwood and W.~Marciano \cite{atw-marc}. This procedure
incorporates leading-log and short-distance loop corrections, and sums
soft-virtual and real-photon corrections to all orders. It does not
incorporate hard-photon bremsstrahlung, which mainly modifies the low
energy portion of the electron spectrum, and is not used in our
analysis. An independent method of studying QED radiative corrections
in semileptonic decays, based on the simulation package PHOTOS
\cite{photos}, has been used to obtain an independent assessment of the
corrections.  The difference between the two methods is used to obtain
the systematic error of this correction. 

\begin{figure}[htbp]
\center{
{\epsfig{figure=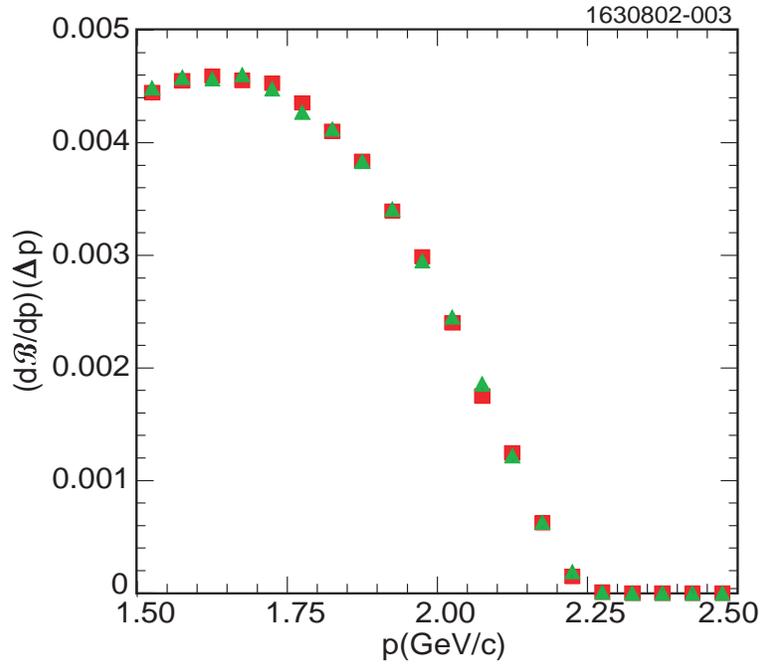,width=4.in,height=3.5in}}}
\caption{\small {Corrected electron (triangles) and muon
(squares) momentum spectra in the
$B$-meson rest frame,
 where $d{\cal B}$ represents the differential semileptonic branching fraction
in the bin $\Delta p$, divided
by the number of $B$ mesons in the sample.}}

\label{bothspcatw}
\end{figure}

Finally, we use a Monte Carlo sample of $b\rightarrow c \ell \bar{\nu}$
events to derive a
matrix to unfold the corrected spectra from the laboratory frame into the $B$-meson rest frame. ($B$
mesons produced at the $\Upsilon (4\rm S)$ by the CESR~ \epm\ collider typically have a momentum of $p_B\sim$ 300 MeV/$c$ in the
laboratory frame.) 
Our lower momentum limit of 1.35 GeV/$c$ for the measurement of the lepton
spectra ensures that end effects in the unfolding procedure do not
introduce distortions into the determination of the spectral moments. The measured spectrum 
includes leptons from $b \rightarrow c \ell \bar{\nu}$ and  $b\rightarrow u \ell \bar{\nu}$. 
Figure \ref{bothspcatw} shows the resulting electron and muon spectra. While the curves shown combine both signs of lepton charges, we have also studied  
positive and negative leptons separately and found good agreement between them. The 
$b\rightarrow u \ell \bar{\nu}$ tail beyond the endpoint of charmed semileptonic decay is
not shown in Fig.~\ref{bothspcatw} but is 
unfolded and added separately to the measured moments.

Our first step is the determination of the truncated moments R$_0$ and R$_1$
defined in Eqs.~(\ref{r2}) and~ (\ref{r1}), respectively. Using the
measured spectra, we evaluate the relevant integrals and obtain
the results shown in Table~\ref{rmom}, where the first error is
statistical and the second is systematic in each quoted number.
Table \ref{err:r01} summarizes our studies of several sources of
systematic uncertainty and their effect on the moments $R_0$ and
$R_1$. The dominant uncertainty for both lepton species is related
to particle identification efficiency. As the moments are ratios
of measured quantities, the effects of several uncertainties,
which are nearly independent  of the lepton energy, cancel. The
overall systematic uncertainties are 0.28\% for R$^{exp}_0$ and
0.06\% for R$^{exp}_1$ for the \elep\ sample, and 0.32\% and
0.06\% for the \mulep\ sample. These are comparable to the
statistical uncertainties.
\begin{table}[htbp]
\begin{center}
\caption{\label{rmom} Measured truncated lepton moments for \elep\, \mulep\ and
combined (weighted average of \elep\ and \mulep).}
\begin{tabular}{l|c|c}
\hline\hline

         &   R$^{exp}_0$                 &   R$^{exp}_1 (\rm GeV)$                 \\ \hline
\elep\   &$0.6184 \pm 0.0016 \pm 0.0017$ &$1.7817 \pm 0.0008 \pm
0.0010$ \\  \mulep\  &$0.6189 \pm 0.0023 \pm 0.0020$ &$1.7802 \pm
0.0011 \pm 0.0011$ \\ \hline $\ell ^{\pm}$ &$0.6187 \pm 0.0014 \pm
0.0016$ &$1.7810 \pm 0.0007 \pm 0.0009$ \\ \hline\hline
\end{tabular}
\end{center}
\end{table}
Since the two moments are extracted from the same spectra, we must use the covariance matrix
E$_{R_0 R_1}$ to extract the HQE parameters. Table \ref{correlation} shows
the numerical values of the E$_{R_0 R_1}$ elements for electrons and muons.
\begin{table}[htpb]
\begin{center}
\caption{\label{err:r01} Summary of the statistical, and systematic
errors on the moments $\mbox{R}^{exp}_0$ and
$\mbox{R}^{exp}_1$.}
\begin{tabular}{l|cc|cc} \hline\hline
  & \multicolumn{2}{c|}{$\delta R_0$($\times 10^3$)}
  & \multicolumn{2}{c}{$\delta R_1$(GeV)($\times 10^3$)}  \\\hline 
  & $e^{\pm}$&$\mu^{\pm}$ & $e^{\pm}$&$\mu^{\pm}$ \\ 
Statistical error & $1.6$ & $2.3$  & $0.8$ & $1.1$  \\
\hline
Continuum subtraction & $0.42$ & $0.30$  & $0.36$ & $0.27$  \\
$J/\psi$ veto           & $0.15$ & $0.07$  & $0.10$ &
$0.08$  \\
$\pi^0$ veto & $0.04$ & N/A & $0.01$ &  N/A \\
Leptons from $b\rightarrow c \rightarrow s(d) \ell \nu$ &
$0.64$ & $0.70$ & $0.20$ & $0.30$ \\
Leptons from $B\rightarrow \tau\ X $ & $0.22$ & $0.25$ & $0.10$ & $0.10$  \\
Fake leptons & $0.04$ & $0.02$ & $0.02$& $0.19$  \\ 
Detection efficiency    & $0.10$ & $0.08$ & $0.20$ & $0.08$ \\
Particle identification efficiency & $0.91$ & $1.52$ & $0.40$ & $0.65$  \\
Electroweak radiative correction & $0.75$ &  $0.43$ &$0.25$  & $0.15$  \\ 
B $\rightarrow X_u\ell \bar{\nu}$  shape uncertainty&$0.50$&$0.40$ & $0.40$ & $0.30$ \\
Unfolding effect & $0.34$ & $0.44$ & $0.14$ & $0.12$ \\
Absolute momentum scale uncert.&$0.70$& $0.70$&$0.50$& $0.50$ \\ 
Total systematic uncertainties  & $1.7 $ & $2.0$ & $1.0$  & $1.1$ \\ 
\hline\hline
\end{tabular}
\end{center}
\end{table}
\begin{table}[htpb]
\begin{center}
\caption{Covariance matrices for the experimental errors on R$^{exp}_0$
and R$^{exp}_1$ moments.} \label{correlation}
\begin{tabular}{l|c} \hline\hline
                & E$_{R_0R_1} (\times 10^6)$ \\ \hline
\elep\  &
$\left( \begin{array}{cc}
5.5 & 1.1 \\
1.1 & 1.6 
\end{array} \right)$

\\ \hline

\mulep\    &

$\left( \begin{array}{cc}
5.2  & 2.2  \\
2.2  & 9.3 
\end{array} \right)$
\\ \hline 

$\ell ^{\pm}$  &

$\left( \begin{array}{cc}
4.5  & 0.8  \\
0.8  & 1.3
\end{array} \right)$
\\ \hline\hline

\end{tabular}

\end{center}
\end{table}

\section{Determination of the HQE parameters}
First, we determine $\bar{\Lambda}$ and $\lambda_1$ using the
published expressions for the moments R$_0$ and R$_1$ in terms of
the HQE parameters\cite{ligeti}. In addition, we  explore the
implications of other constraints derived from the lepton energy
spectrum \cite{chris}.

\subsection{Determination of $\bar{\Lambda}$ and $\lambda _1$ from
the moments R$_0$ and R$_1$.}

The theoretical expressions\cite{ligeti,chris} relating the
spectral moments to the HQE parameters include correction terms 
accounting for electroweak radiative effects and the
unfolding from the laboratory to the rest frame. We do not use
these additional terms because our data are corrected for these
effects. The  non-perturbative expansion
\cite{manohar-wise,bigi,gremm-kap} includes terms through order
$1/M_B^3$ .

\begin{figure}[ht]
\center{\epsfig{figure=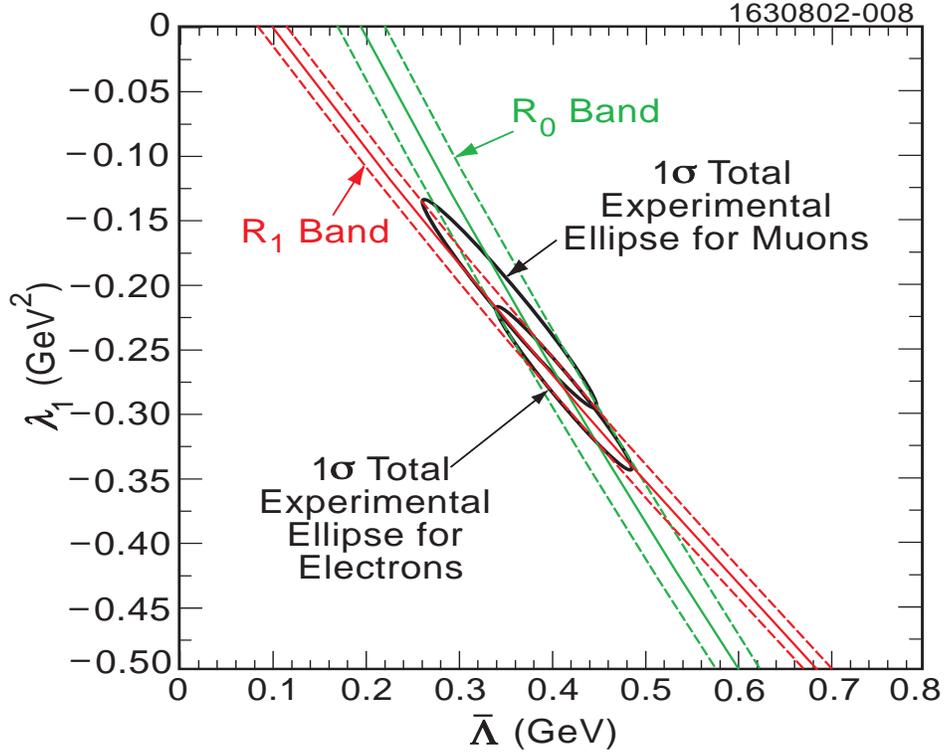,width=5.0in,height=4in}}
\caption{ Constraints on the HQE parameters $\lambda_1$ and
$\bar\Lambda$ from our measured moments of the electron spectrum
R$_0$ and R$_1$. The contours represent $\Delta\chi^2=1$ for the
combined statistical and systematic errors on the measured values.
The parameters $\lambda_1$ and $\bar \Lambda$ are computed in the
$\overline{MS}$ scheme to order $1/M^3_B$ and $\beta_0
\alpha_s^2$. } \label{elpsedata}
\end{figure}

The values of the HQE parameters and their
experimental uncertainties are obtained by calculating the $\chi^2$
from the measured moments R$^{exp}_0$ and R$^{exp}_1$ and the
covariance matrix $\mbox{E}_{R_0 R_1}$
\begin{equation}
\chi^2= \sum_{\alpha=0}^{\alpha=1} \sum_{\beta=0}^{\beta=1}
(\mbox{R}^{exp}_{\alpha}-\mbox{R}^{th}_{\alpha}) \; \;
\mbox{E}^{-1}_{R_0R_1}  \; \;
(\mbox{R}^{exp}_{\beta}-\mbox{R}^{th}_{\beta}),
\end{equation}
\noindent where R$^{th}_{0}$ and R$^{th}_{1}$ are

\begin{eqnarray}
{\mbox{R}^{th}_{0}} &=& 0.6581 - 0.315 \left(\frac{{\bar{\Lambda}}}{{\bar{M}_B}}\right)
- 0.68 \left(\frac{{\bar{\Lambda}}}{{\bar{M}_B}}\right)^2 -1.65
\left(\frac{\lambda_1}{{\bar{M}_B}^2}\right)- 4.94
\left(\frac{\lambda_2}{{\bar{M}_B}^2}\right)   \nonumber \\
  & &+ \left|\frac{V_{ub}}{V_{cb}}\right|^2 \left(0.87 - 
3.8 \frac{{\bar{\Lambda}}}{{\bar{M}_B}}\right)
- 1.5 \left(\frac{{\bar{\Lambda}}}{{\bar{M}_B}}\right)^3 - 7.1
\left(\frac{{\bar{\Lambda}} \lambda_1}{{\bar{M}_B}^3}\right)
- 17.1 \left(\frac{{\bar{\Lambda}} \lambda_2}{{\bar{M}_B}^3}\right) \nonumber \\
 & & -1.8 \left(\frac{\rho_1}{{\bar{M}_B}^3}\right) +
 2.3 \left(\frac{\rho_2}{{\bar{M}_B}^3}\right)
- 2.9 \left(\frac{\tau_1}{{\bar{M}_B}^3}\right) - 1.5 \left(\frac{\tau_2}{{\bar{M}_B}^3}
\right)
- 4.0 \left(\frac{\tau_3}{{\bar{M}_B}^3}\right) -
4.9 \left(\frac{\tau_4}{{\bar{M}_B}^3}\right)    \nonumber \\
 & & - \frac{\alpha_s}{\pi} \left(0.039 + 0.18\frac{{\bar{\Lambda}}}{{\bar{M}_B}}\right) -
0.098 \left(\frac{\alpha_s}{\pi}\right)^{2} \beta_{0}
\label{lig0}
\end{eqnarray}
and
\begin{eqnarray}
{\mbox{R}^{th}_{1}} &=& 1.8059 - 0.309 \left(\frac{{\bar{\Lambda}}}{{\bar{M}_B}}\right) 
- 0.35
\left(\frac{{\bar{\Lambda}}}{{\bar{M}_B}}\right)^2 - 
2.32 \left(\frac{\lambda_1}{{\bar{M}_B}^2}\right)
- 3.96 \left(\frac{\lambda_2}{{\bar{M}_B}^2}\right) \nonumber \\
  & &+\left|\frac{V_{ub}}{V_{cb}}\right|^2 \left(1.33-10.3 \frac{{\bar{\Lambda}}}{{\bar{M}_B}}\right)
- 0.4 (\frac{{\bar{\Lambda}}}{{\bar{M}_B}})^3 - 5.7 (\frac{{\bar{\Lambda}}
\lambda_1}{{\bar{M}_B}^3}) - 6.8\left(\frac{{\bar{\Lambda}} \lambda_2}
{{\bar{M}_B}^3}\right) \nonumber \\
 & & - 7.7 \left(\frac{\rho_1}{{\bar{M}_B}^3}\right)-1.3 \left(\frac{\rho_2}{{\bar{M}_B}^3}\right)
- 3.2 \left(\frac{\tau_1}{{\bar{M}_B}^3})- 4.5 (\frac{\tau_2}{{\bar{M}_B}^3}\right)
- 3.1 \left(\frac{\tau_3}{{\bar{M}_B}^3}\right)- 4.0
\left(\frac{\tau_4}{{\bar{M}_B}^3}\right) \nonumber \\
 & & - \frac{\alpha_s}{\pi} \left(0.035 + 0.07
\frac{{\bar{\Lambda}}}{{\bar{M}_B}}\right) - 0.082 \left(\frac{\alpha_s}{\pi}\right)^{2} \beta_{0}.
\label{lig1}
\end{eqnarray}
In Fig.~\ref{elpsedata} we show the $\Delta\chi^2=1$ contours for
electrons and muons corresponding to the quoted experimental
uncertainties.

\begin{figure}[ht]
\center{\epsfig{figure=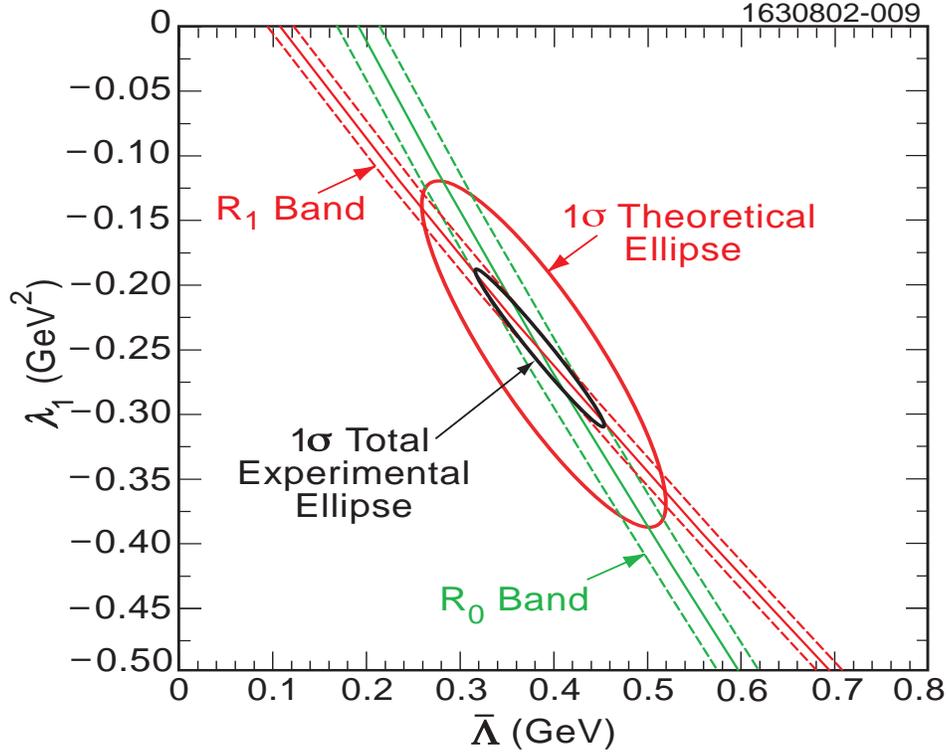,width=5.in,height=4.in}}
\caption{The constraints from our combined electron and muon R$_0$ and
R$_1$ moments, with $\Delta\chi^2=1$ contours for total experimental
and theoretical uncertainties. The parameters $\lambda_1$ and
$\bar \Lambda$ are computed in the $\overline{MS}$ scheme to order
$1/M^3_B$ and $\beta_0 \alpha_s^2$.}
\label{therrepse}
\end{figure}

The theoretical uncertainties on the HQE parameters are determined
by varying, with flat distributions, the input parameters within
their respective errors: $|{V_{ub}}/{V_{cb}}|= 0.09 \pm 0.02$ 
\cite{adam1},
$\alpha_s = 0.22 \pm 0.027$, $\lambda_{2} = (0.128\pm 0.010)\ 
\rm GeV ^2$\cite{gremm-kap},  $\rho_2 = 0 \pm (0.5)^3$ GeV$^3$, and  $\tau_i = 0.0 \pm
(0.5)^3$ GeV$^3$ \cite{gremm-kap}.  The parameter $\rho_1$ is taken as
 $0.5(0.5)^3 \pm 0.5(0.5)^3$
GeV$^3$, because it is expected to be positive \cite{ikaros-rho1}. The contour that contains 68\% of the
probability is shown in Fig.~\ref{therrepse}. This procedure for
evaluating the theoretical uncertainty from the unknown expansion
parameters that enter at order $1/M_B^3$ is similar to that used
by Gremm and Kapustin \cite{gremm-kap}  and Bauer and Trott
\cite{chris}, but different from the procedure used in our
analysis of hadronic mass moments \cite{eht-moments}. The dominant
theoretical uncertainty is related to the $1/M_B^3$ terms in the
non-perturbative expansion discussed before. Reference
\cite{chrisligeti} has explored the convergence of the
perturbative and non-perturbative series appearing in the
expressions for the moments described in this paper. The most
conservative estimate gives a truncation error of at most 20\% of the central
values of the HQE parameters.
The theoretical uncertainties presented in this paper do not
include this truncation error.
The measured ${{\lambda}_{1}}$ and $\bar{\Lambda}$ are given in
Table \ref{statsyst}.

\begin{table}[h]
\begin{center}
\caption{Measured ${\lambda}_{1}$ and $\bar{\Lambda}$ values,
including statistical, systematic, and theoretical errors.}
\label{statsyst}
\begin{tabular}{l|c|c} \hline\hline

                & ${\lambda}_{1}$(GeV$^2$)  & $\bar{\Lambda}$(GeV)  \\ \hline
\elep &$-0.28\pm0.03_{stat}\pm0.06_{sys}\pm0.14_{th}$
&$0.41\pm 0.04_{stat}\pm 0.06_{sys}\pm0.12_{th}$\\ \hline
\mulep    &$-0.22\pm0.04_{stat}\pm0.07_{sys}\pm0.14_{th}$
&$0.36\pm 0.06_{stat}\pm 0.08_{sys}\pm0.12_{th}$\\ \hline 

$\ell ^{\pm}$ &$-0.25\pm0.02_{stat}\pm0.05_{sys}\pm0.14_{th}$
&$0.39\pm 0.03_{stat}\pm 0.06_{sys}\pm0.12_{th}$\\ \hline\hline

\end{tabular}

\end{center}
\end{table}

\begin{figure}[ht]
\center{\epsfig{figure=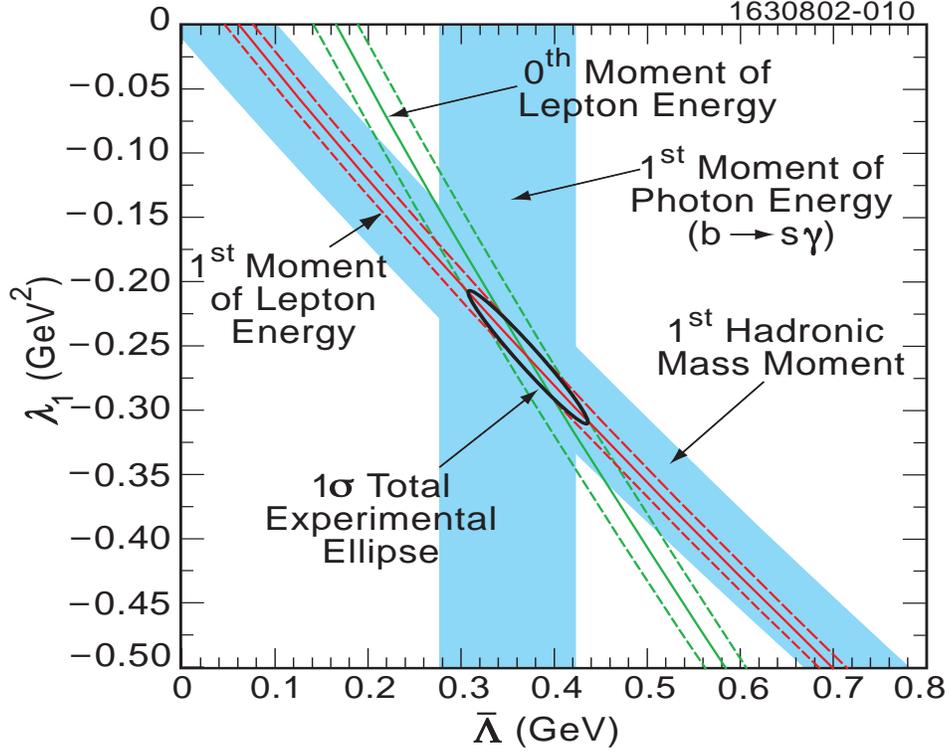,width=5.in,height=4.in}}
\caption{Experimental constraints from the ${\bar B}\to X \ell{\bar\nu}$ hadronic
mass moment and $b\rightarrow s \gamma$ $E_\gamma$ moment
\cite{eht-moments} compared with the combined electron and muon R$_0$
and R$_1$ constraints.
The parameters $\lambda_1$ and $\bar \Lambda$ are
computed in the $\overline{MS}$ scheme to order $1/M^3_B$ and $\beta_0
\alpha_s^2$.}
\label{hadbsglep}
\end{figure}

A previous CLEO measurement used the first moment of the hadronic
recoil mass \cite{eht-moments}
and the first photon energy moment from the $b \rightarrow s \gamma$
process \cite{bsgamma}. Fig.~\ref{hadbsglep} shows a comparison of our results
with the previously published ones. We overlay the experimental ellipse
from the electron
and muon combined spectral
measurement, using in this case $|\frac{V_{ub}}{V_{cb}}|= 0.07$ to be 
consistent with the
assumptions in that paper. The agreement is good, although the theoretical uncertainties do not warrant a very
precise comparison.

Using the expression for the full semileptonic decay width given
in Eq.~(\ref{vcbf}), we can extract $|{\mbox{V}}_{cb}|$. We use
$\Gamma_{sl}^{exp}=(0.43 \pm 0.01){\times} 10^{-10}$MeV
\cite{pdg2002}.  Assuming the validity of quark-hadron duality, we
obtain

\begin{equation}
|{\mbox{V}}_{cb}|=(40.8\pm 0.5 _{\Gamma_{sl}}\pm 0.4 _{\lambda_1,
\bar{\Lambda}}\pm 0.9 _{th})\times 10^{-3},
\end{equation}
where the first uncertainty is from the experimental
value of the semileptonic width, the second uncertainty is from the
HQE parameters (${\lambda}_{1}$ and $\bar{\Lambda}$), and the third
uncertainty is the theoretical uncertainty obtained as described above.

\subsection{Determination of the short range mass $m_b^{1\rm S}$.}

\begin{figure}[ht]
\center{\epsfig{figure=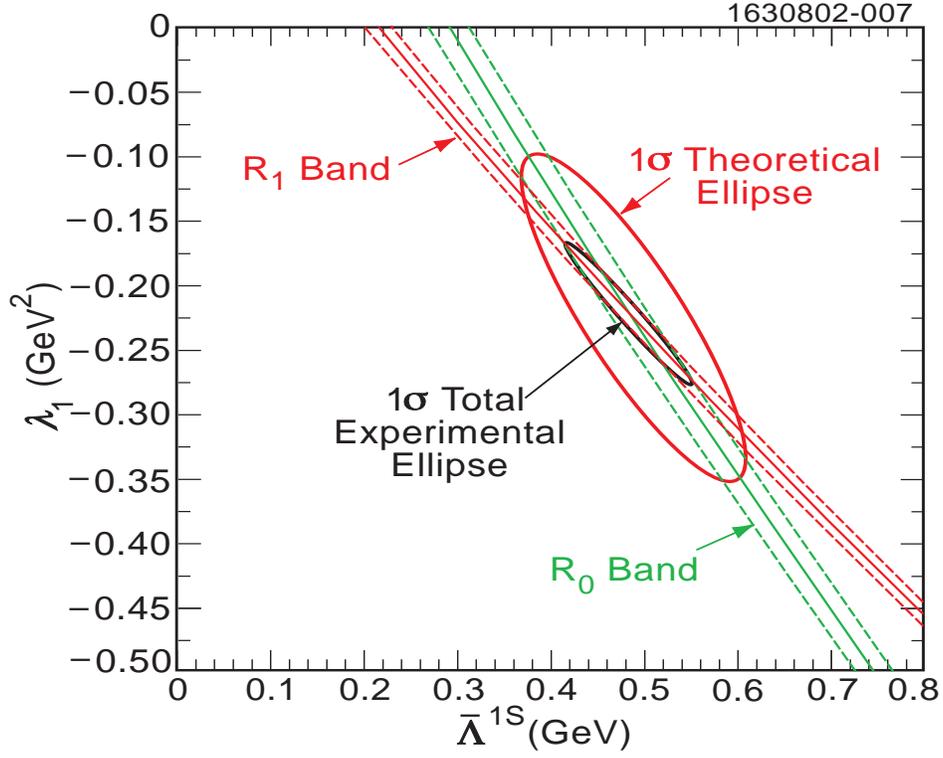,width=5.in,height=4.in}}
\caption{The combined electron and muon R$_0$ and R$_1$
constraints on the parameters $\bar\Lambda ^{1\rm S}$ and $\lambda _1$, 
showing the $\Delta\chi^2=1$ contours for total experimental
and theoretical uncertainties, using the constraints in
Ref.~\cite{chris}.} \label{lamb1s:fig}
\end{figure}

We use the formalism of Ref. \cite{chris} to extract the short
range mass of the $b$ quark \mus , defined as $m_b^{1\rm S}\equiv
{\bar M}_B- {\bar{\Lambda}}^{1 \rm S}$. Table \ref{lamb1s}
summarizes the measurement of $\bar{\Lambda}^{1\rm S}$ and
$\lambda _1$ for electrons and muons separately, and for
the combined sample. 
Fig.~\ref{lamb1s:fig} shows the corresponding bands and the $\delta \chi^2 =1$
contour. The theoretical uncertainty is extracted using the method
described above. Our result, $m_b^{1\rm S}=(4.82\pm 0.07_{exp}\pm
0.11_{th})\ {\rm GeV}/c^2$, is in good agreement with a previous
estimate of $m_b^{1\rm S}$ \cite{h1} derived from $\Upsilon (1\rm
S)$ data, \mus$=4.69 \pm 0.03$ GeV/$c^2$.

\begin{table}[h]
\begin{center}
\caption{The measured ${\bar{\Lambda}}^{1\rm S}$ and $m_b^{1\rm
S}$. The quoted errors reflect statistical, systematic, and
theoretical uncertainties, respectively.} \label{lamb1s}
\begin{tabular}{l|c|c|c} \hline\hline
           &${\bar{\Lambda}}^{1\rm S}$(GeV)
                       &$m_b^{1\rm S}$(GeV/$c^2$)& $\lambda _1\rm (GeV ^2)$ \\ 
\hline
\elep\ &$0.52\pm0.04_{stat}\pm0.06_{sys}\pm0.11_{th}$
                       & $ 4.79 \pm 0.07_{exp}\pm 0.11_{th}$ & -0.26$\pm 0.03_{stat}\pm 0.05_{sys}\pm 0.12_{th}$\\ 
\mulep\    &$0.46\pm0.05_{stat}\pm0.08_{sys}\pm0.11_{th}$
                       & $ 4.85 \pm 0.0_{exp}\pm 0.11_{th}$
& -0.19$\pm 0.04_{stat}\pm 0.07_{sys}\pm 0.12_{th}$\\ 
$\ell ^{\pm}$ & $0.49 \pm 0.03_{stat} \pm 0.06_{sys} \pm 0.11 _{th}$ &
4.82 $\pm 0.07 _{exp}\pm 0.11 _{th}$ & -0.23$\pm 0.02_{stat}\pm 0.05_{sys}\pm 0.12_{th}$\\
\hline\hline
\end{tabular}

\end{center}
\end{table}

\subsection{Measurements of additional spectral moments and
implications for the HQE parameters}

\begin{figure}[ht]
\center{\epsfig{figure=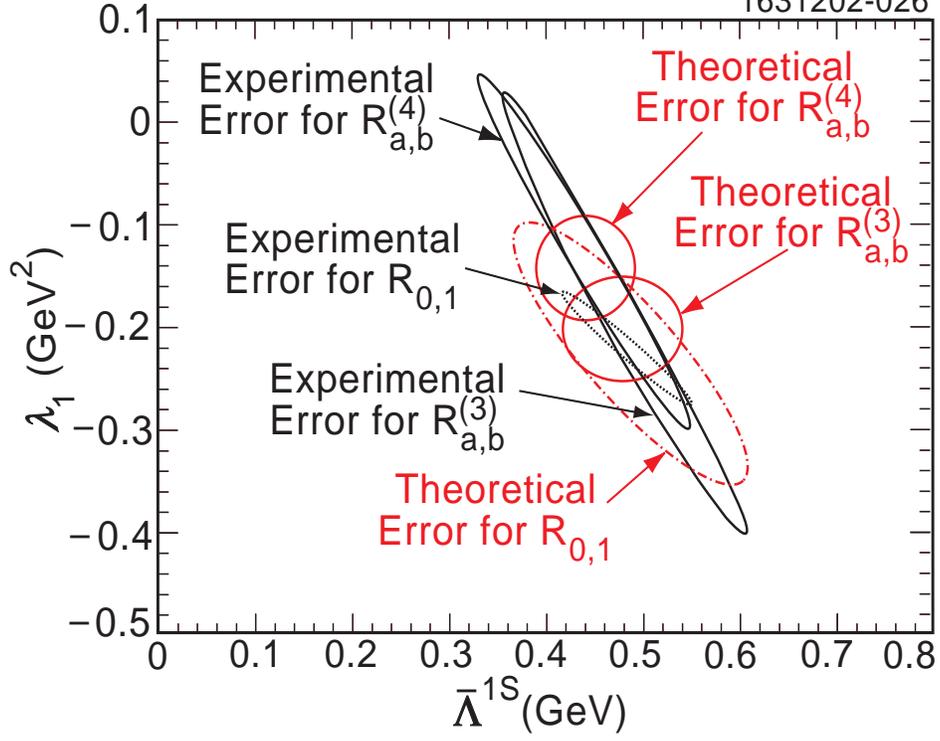,width=5.0in,height=4in}}
\caption{ Constraints on the HQE parameters $\lambda_1$ and
${\bar{\Lambda}}^{1\rm S}$ from all our measured spectral
moments.} \label{all1s}
\end{figure}

We apply the same experimental procedure described before to
measure a variety of spectral moments. In particular, we measure the moments $R^{(3)}_{a}$,
$R^{(3)}_{b}$, $R^{(4)}_{a}$, and $R^{(4)}_{b}$ defined in Eqs.~(\ref{r3a})-(\ref{r4b}).
Tables \ref{r3ab} and \ref{r4ab} summarize their measured values,
as well as the statistical and systematic errors. Fig.~\ref{all1s}
shows the measured $\bar{\Lambda}^{1\rm S}$ and $\lambda _1$ with
these two sets of constraints, as well as the constraints derived
from the moments $R_0$ and $R_1$. Although we are able to confirm
that $1/M_B^3$ terms produce much smaller uncertainties using $R^{(3,4)}_{a,b}$, the experimental errors are larger in this case because of the similar slopes for the two
constraints. The uncertainty ellipses are still sizeable, but
the systematic and theoretical uncertainties have a different nature
and magnitude and thus the overall
agreement is significant.

\begin{table}[htpb]
\begin{center}
\caption{\label{r3ab} Measured truncated lepton moments $R^{(3)}_{a,b}$
for \elep ,
\mulep, and their weighted average.}
\begin{tabular}{l|c|c}
\hline
         & R$^{(3)}_a$(GeV$^{-1.3}$)  & R$^{(3)}_b$(GeV$^{0.9}$)   \\ \hline\hline
\elep   &$0.3013 \pm 0.0006_{stat} \pm 0.0005_{sys}$
&$2.2632 \pm 0.0029_{stat} \pm 0.0026_{sys}$ \\ 

\mulep\  &$0.3019 \pm 0.0009_{stat} \pm 0.0007_{sys}$
&$2.2611 \pm 0.0042_{stat} \pm 0.0020_{sys}$ \\ 

$\ell ^{\pm}$ &$0.3016 \pm 0.0005_{stat} \pm 0.0005_{sys}$
&$2.2621 \pm 0.0025_{stat} \pm 0.0019_{sys}$ \\ \hline\hline
\end{tabular}
\end{center}

\end{table}

\begin{table}[htpb]
\begin{center}
\caption{\label{r4ab} Measured truncated $R^{(4)}_{a,b}$ moments for \elep , 
\mulep, and their weighted average.}
\begin{tabular}{l|c|c}
\hline\hline
         & R$^{(4)}_{a}$(GeV$^{0.8}$)  & R$^{(4)}_{b}$(GeV$^{-0.4}$)   \\ \hline
\elep\   &$2.1294 \pm 0.0028_{stat} \pm 0.0027_{sys}$
&$0.6831 \pm 0.0005_{stat} \pm 0.0007_{sys}$ \\ \hline

\mulep\  &$2.1276 \pm 0.0040_{stat} \pm 0.0015_{sys}$
&$0.6836 \pm 0.0008_{stat} \pm 0.0014_{sys}$ \\ \hline 

$\ell ^{\pm}$ &$2.1285 \pm 0.0024_{stat} \pm 0.0018_{sys}$
&$0.6833 \pm 0.0005_{stat} \pm 0.0006_{sys}$ \\ \hline\hline
\end{tabular}
\end{center}
\end{table}

Finally, we extract the duality moments $D_3$, and $D_4$ from the
measured shape of the electron and muon spectra.  The theoretical predictions for these 
moments in Ref.~\cite{chris}, evaluated using the
values of $\bar{\Lambda}^{1\rm S}$ and $\lambda _1$ reported in
this paper, are compared with the measured $D_{3,4}$ from the
combined lepton sample in Table \ref{dualth}. The agreement is
excellent and thus no internal inconsistency of the theory is
uncovered in this analysis.
\begin{table}[htpb]
\begin{center}
\caption{Measured duality moments and theoretical predictions using 
the values $\lambda_1$ and $\bar{\Lambda}^{1\rm S}$ reported in this paper. The errors reflect the experimental uncertainties in these parameters 
and the theoretical errors, respectively.}
\label{dualth}
\begin{tabular}{l|c|c} \hline\hline

       & Experimental           &Theoretical \\ \hline
D$_{3}$&$0.5193\pm0.0008_{exp}$& $0.5195\pm 0.0006_{\lambda_1,\bar{\Lambda}^{1\rm S}}\pm 0.0003_{th}$\\ 
D$_{4}$&$0.6036\pm0.0006_{exp}$& $0.6040\pm 0.0006_{\lambda_1,\bar{\Lambda}^{1\rm S}}\pm 0.0005_{th}$\\ \hline\hline
\end{tabular}
\end{center}
\end{table}

\section{Conclusion}

We have measured the lepton momentum spectra in ${\bar
B} \rightarrow X \ell {\bar \nu}$ ($\ell = e$  and $\mu$)
for p $\ge$ 1.5 GeV/$c$ in the B rest frame. From these, we determine the
spectral moments R$_0$, R$_1$, R$^{(3)}_{a}$, R$^{(3)}_{b}$, R$^{(4)}_{a}$,
R$^{(4)}_{b}$, D$_{3}$ and D$_{4}$.

Using the moments R$_0$ and R$_1$ we extract the HQE parameters
$\bar{\Lambda}= (0.39 \pm 0.03_{stat} \pm 0.06_{sys} \pm
0.12_{th}$) GeV and ${\lambda}_{1}= (-0.25 \pm 0.02_{stat} \pm
0.05_{sys}\pm 0.14_{th}$) GeV$^2$. These results imply that the
pole mass $m_b = (4.90 \pm 0.08_{exp}\pm 0.13_{th}$) GeV/$c^2$.
The short range mass $m_b^{1\rm S}$ is found to be ($4.82 \pm
0.07_{exp}\pm 0.11_{th}$) GeV/$c^2$. We obtain
$|{\mbox{V}}_{cb}|=(40.8\pm 0.5_{\Gamma_{sl}}\pm 0.4_{\lambda_1,
\bar{\Lambda}}\pm 0.9_{th})\times 10^{-3}$.

Finally, an extensive study of different spectral moments shows
good agreement between independent determinations of the HQE
parameters.
\section{Acknowledgements}
We gratefully acknowledge the effort of the CESR staff 
in providing us with excellent luminosity and running conditions.
M. Selen thanks the Research Corporation, 
and A.H. Mahmood thanks the Texas Advanced Research Program.
This work was supported by the National Science Foundation 
and the U.S. Department of Energy.

We would also like to thank C.~Bauer, Z.~Ligeti, and N.~G.~Uraltsev for
helpful discussions.

\newpage

\end{document}